\documentclass[preprint,floats,aps]{revtex4}
\usepackage{bm}
\usepackage{graphicx}
\begin{document}
\topmargin=-0.3cm
\title{Kinetic description of hadron-hadron collisions}
\author{ Zhi Guang Tan$^{1-2}$, S.Terranova$^{1}$
and A. Bonasera$^{1,3}$
 \footnote{Email: bonasera@lns.infn.it}}
\affiliation{
$^1$  Laboratorio Nazionale del Sud, Istituto Nazionale Di Fisica Nucleare,
      Via S. Sofia 44, I-95123 Catania, Italy \\
$^2$  Institute of Particle Physics, Huazhong Normal University, Wuhan,
      430079 China\\
$^3$ Libera Universita'  Kore, via della Cittadella 1, I-94100 Enna, Italy      
}
\begin{abstract}
A transport model  based on the mean free path approach  to describe pp collisions is proposed.  We assume that hadrons can be treated as  bags of partons similarly to the  MIT bag model. When the energy density in the collision is higher than a critical value, the bags break and partons are liberated.  The partons expand and can make coalescence to form new hadrons.  The results obtained compare very well with available data and some prediction for higher energies collisions are discussed.  Based on the model we suggest that a QGP could
already be formed in the pp collisions at high energies. 
\end{abstract}
\maketitle

In  recent works, we have proposed a kinetic approach to deal with a hot pion gas and a possible phase transition to the Quark Gluon Plasma (QGP)\cite{tan}.  In our approach we treated the hadrons as bags of partons similar to the MIT bag model: if in a particle-particle collision the energy density is higher than a critical value obtained from the bag constant,
new partons are created.  Those partons can evolve in time and collide again with other particles.  If the partons travel
 in a region whose density is below a critical value, they can coalesce and form new hadrons.  The equation of state at
 zero barionic density was calculated and the effects of a mean field to reproduce 
Lattice QCD (LQCD) results  were implemented as well\cite{tan,karsch}.  Up to date LQCD results suggest that there is a cross over from a meson system to a QGP at a temperature of about 175 MeV \cite{karsch}.  

It is the purpose of this work to extend the model to pp (and later to AA)  collisions which are some of the most powerful tools to 
detect experimentally a transition to the QGP.
Of course exact  microscopic  simulations for out of equilibrium-finite systems are out of reach at present.  On the other hand transport approaches  \cite{gei} have been very useful in the past in describing many features of lower energies heavy ion collisions.  Generalizations to relativistic energies of low energy heavy ion collisions \cite{cug,bert,sa1} (known as  Boltzmann Uehling Uehlenbeck (BUU), Vlasov(VUU)/ Landau (LV)) have been proposed.
The method we discuss in
this work is known as Boltzmann Nordheim Vlasov (BNV) approach at low energies\cite{aldo,daimei,tan}.  It
is based on the concept of the mean free path approach\cite{aldo}.

  We can easily include the possibility of a QGP using the bag model \cite{tan,las,wong}. In fact, for each elementary hadron-hadron collision, we can calculate the local energy density and the pressure.  If such quantities overcome the bag pressure and energy density, then $n_q=n_{\bar q}$ quarks and antiquarks and $n_g$ gluons are created.  The number of quarks and gluons are calculated assuming local thermal equilibrium.  In this way we can simulate a hadron gas and its transition to the QGP.  In \cite{tan},
  we discussed  the cases of $N_f=0,2,3$, where $N_f$ is the number of flavors plus a mean field. We also discussed the possibility that quarks could recombine and gluons decay into two quarks 
  during the dynamical process.   We saw that in order to have a reasonable description of hadronization the local quark (gluon) density must not exceed a critical density calculated from the MIT bag model.  If the local density is larger than such a 
  value (which is equivalent to having a high temperature or energy density) the quarks cannot recombine to form hadrons (or the gluons decay into $q \bar q$ pair).  From a comparison of our results to LQCD we realized that in the QGP phase our $\epsilon/T^4$ is higher than the value suggested by
  lattice calculation and indeed it approaches the Stefan-Boltzmann limit as it should be.  Knowing that there is Debye color screening in the QGP phase we calculated the corresponding  mean field and adjusted the parameters to reproduce lattice data\cite{tan}.  
  
  In order to use the same approach for out of equilibrium situations we have to deal first with
energetic pp (or ee) collisions.  This is the purpose of our paper and we will show in the following that keeping the main ingredients of the model but assuming a suitable momentum distribution of the partons after they are created
 we can reproduce reasonably well some data on pp collisions.  Of course, the EOS  which can be obtained from
 the model remains unchanged by these modifications.  Thus in our approach we can describe with some accuracy,
 both elementary collisions as well as collective properties of the systems.      Quantum statistics (i.e. Pauli and Bose statistics) are  included
 similarly to   \cite{bert1}
 for Bose and \cite{sa04} for Fermi statistics\cite{tan}.

 The number of particles that are produced in each elementary collision increases with increasing beam energy in agreement to data.  We make some prediction for pp collisions at 5 TeV which will be soon studied experimentally at CERN.   
 The model assumes that hadrons can be produced from partons coalescence similar to the results of low energy heavy
 ion collisions to study a liquid-gas phase transition\cite{hagel}.  We analyze our results in terms of the Fisher model
 for a liquid-gas phase transition and surprisingly we find a  power law mass distribution with an exponent 
 $\tau=2.3$.  This is an expected value from Fisher's approach\cite{dorso}.  For comparison, we have collected some
 data obtained in Au+Au collisions at RHIC and we found a similar mass distribution.  This calls for  a critical 
 behavior of the system.  We suggest that this could be due to the rapid expansion of the system after the partons
 have been created.  If we push the analogy to the liquid-gas phase transition, we can assume that the vapor is made
 of partons and maybe pions, while the other hadrons make the liquid part.  After the QGP is created, the system 
 expands rapidly and since the partons cannot remain isolated they coalesce.  This results in a rapidly falling (power law)
 mass distribution. 
 
 \section{Formalism}

 The mean free path method discussed above has been studied in detail at low energies and it has been
shown to solve the Boltzmann equation in the cases
 where an analytical solution is known \cite{aldo}.  We have
generalized the approach to keep into account relativistic effects.  The particles move on straight lines
during collisions since we have not implemented any field in this exploratory study.  For short we name the method proposed
 as Relativistic Boltzman equation (ReB).  In order to test our approach, we have discussed also some simple cases where analytical solutions are known to verify the sensitivity of our numerical approximations \cite{aldo,daimei,tan}.

In the present  calculations we  simulate pp collisions at a given energy and impact parameter.  For each energy,
events are obtained changing the initial impact parameter.

\begin{figure}[ht]
\centerline{\hspace{-0.9in}
\includegraphics[width=3.5in,height=2.5in,angle=0]{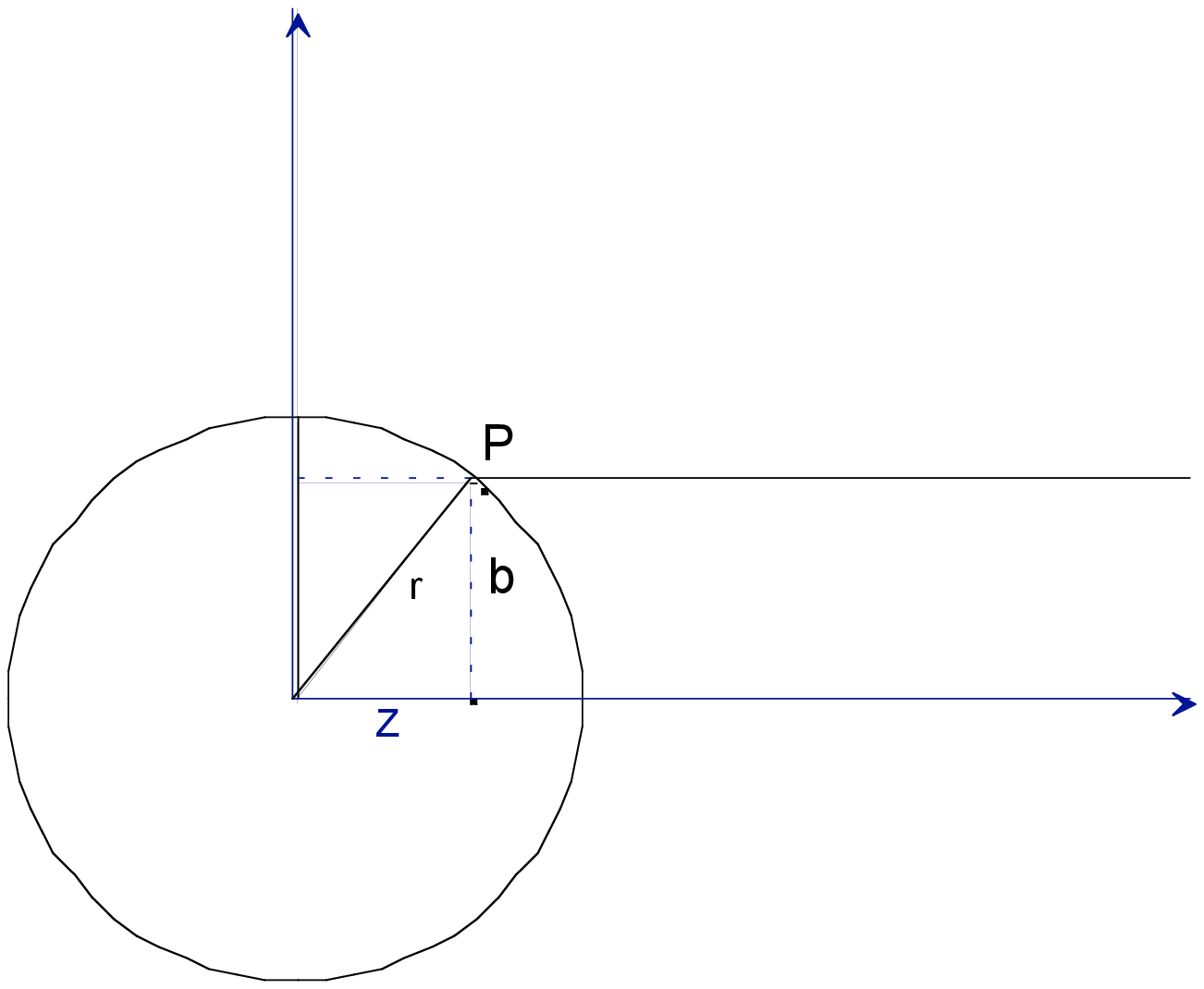}}
\vspace{0.6in} \caption{Schematic view of the pp collision process. b is the impact parameter and r the relative distance between the two protons.  }
\label{fig1}
\end{figure}

We recall how we include partons in our approach.  First,   for a massless quark and gluon plasma in equilibrium the following relations hold for the pressure P, quark (antiquark and gluon) density $n_{q,(\bar q),(g)}$ and energy density $\epsilon$ versus temperature T\cite{las,wong,karsch,tan}:
\begin{equation}
P=g_{tot}\frac{\pi^2}{90}T^4;	
	n_q=n_{\bar q}=1.202\frac{3g_{q}}{4\pi^2} T^3;	
	n_g=1.202 \frac{g_g}{\pi^2}T^3;
\end{equation}

where $\epsilon=3P$, $g_{tot}=16+\frac{21}{2} N_f$, $N_f$ is the number of flavors.  In the MIT bag model\cite{las,wong} quarks and gluons are confined in the bag if the pressure is less than the critical pressure B that the bag can sustain.   Thus from the previous equation we can assume that the quarks and gluons are liberated in a collision if the energy density is larger then 3B.  This gives a critical energy density $\epsilon_c=3B=1.69 GeVfm^{-3}$ using $B^{1/4}=0.242 GeV$\cite{las,wong} and for three flavors.
  For each h-h elementary collision we know  the energy of the  collision and the interaction  volume from the distance between the colliding particles in their center of mass system. This distance is taken as the radius of a sphere enclosing the two colliding particles and subsequently the newly formed partons. Thus we can calculate the number of quarks, antiquarks and gluons as a function of the energy density liberated in the collision inverting equation (1).  These relations can be easily generalized to the cases of finite quark masses. In this work we will use the values  of 5,10 and 160 MeV for u,d and
 s-quark masses respectively.
\begin{figure}[ht]
\centerline{\hspace{-0.5in}
\includegraphics[width=4.5in,height=3.0in,angle=0]{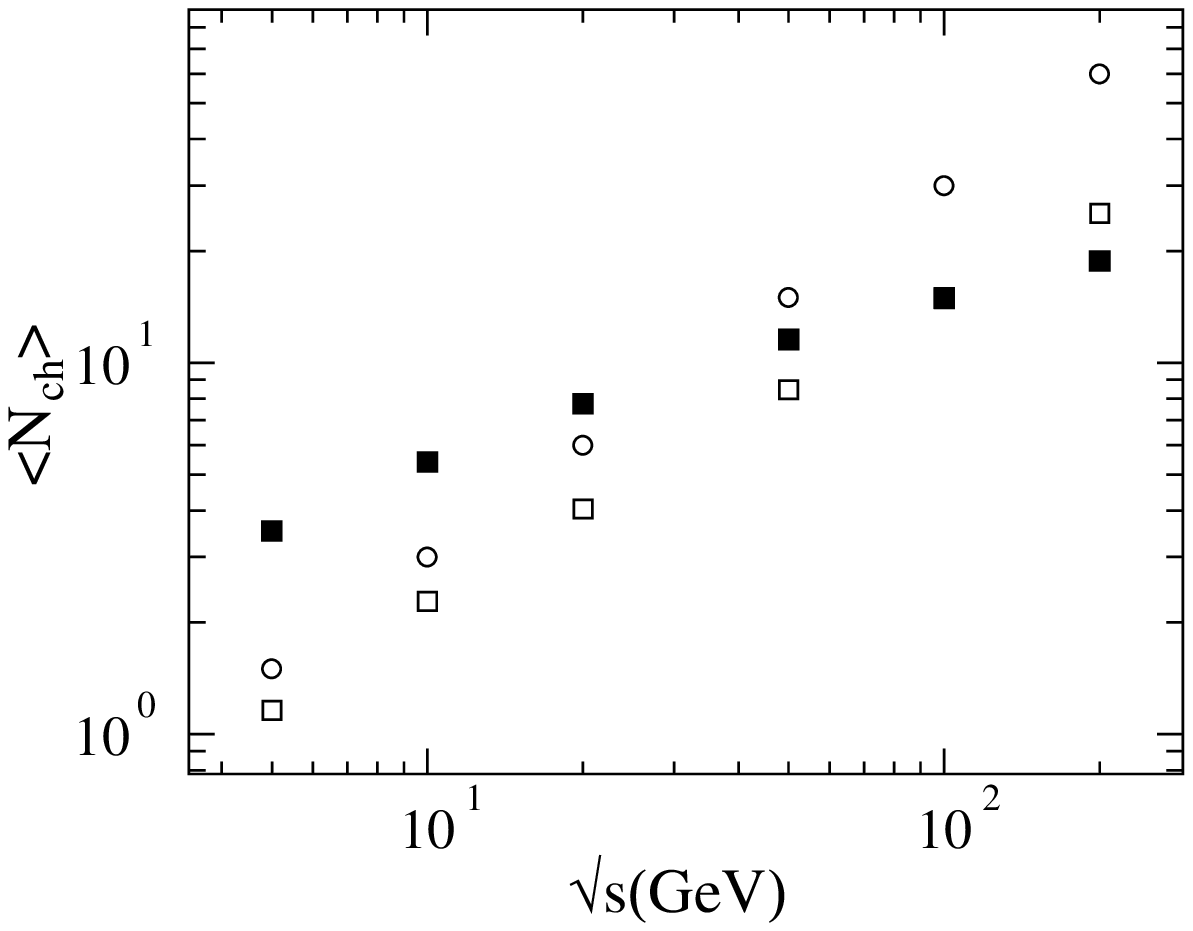}}
\vspace{0.2in} \caption{Analytical estimate of the number of charged  particles produced in pp collisions as function of energy. Data are given by
the full squares, calculations are given by the open circles (without probability corrections) and open squares.   }
\label{fig2}
\end{figure}
We stress that these relations are 
 strictly valid in thermal equilibrium but we are extending them here for non equilibrium cases, i.e. the relevant control parameter is the
 energy density and not the temperature.  Of course, this is
 a strong ansatz and it could be justified if the system is very chaotic, thus many particles should be created for each collision,
 and also after  averaging over many ensembles to have some statistical meaning.  A crucial point of this approach is
 the momentum distribution just after the partons are created.  For particles  in a box the distribution did not really
  matter much, since, because of elastic and inelastic collisions equilibrium is finally reached.  This is not true in pp 
  collisions, in fact we know experimentally that in   such cases equilibrium is not reached.  Thus we assume that
  the partons created have originally a momentum given by a fraction of the initial z-beam momentum.  We assume 
  (as we did in \cite{tan}), that the partons can collide successively elastically and inelastically and we fit the
  corresponding cross sections to reproduce the pseudorapidity experimental data. 
  
    If in a collision between a parton and a hadron the local energy density is larger than the critical value, new partons are liberated from the hadron similarly to the mechanism discussed above. After being created, the partons expand and 
    coalesce into hadrons if the local density become smaller than a critical value obtained from the MIT bag model\cite{tan}. In the model, coalescence occurs  through resonances which decay later on\cite{enzo}.  We have included known
    resonances and their decay using  parametrizations of the data up to   2 GeV.     
  
    Given the essential ingredients of the model, the crucial point now is to reproduce for instance the number of
    produced particles as a function of energy.  We can estimate analytically the number of produced particles making some simplifications.  Let us consider two colliding protons along the z-axis and at impact parameter b, assume the
    particles move on a straight line until they collide.  We can estimate a critical volume  $V_c$ when the energy density is 
  equal to the critical value for breaking the bags.  The situation is depicted in figure (1), and we  easily  get:
   
    \begin{equation}
V_c=\frac{\sqrt(s)}{\epsilon_c}=4/3\pi (\sqrt(z_0^2+b^2))^3;
\label{lv1}
\end{equation}

$z_0$ is the coordinate where, for a given b, the volume enclosing the two colliding particles is equal to the critical one,
see fig.1.
The average (over impact parameters) number of produced quarks is given by:
 \begin{equation}
<N_q>=\frac{\int^{b_{max}}_0 b N(b,\sqrt(s))\Pi(b) db}{\int^{b_{max}}_0 b db};
\label{lv2}
\end{equation}

and the maximum impact parameter $b_{max}$ can be obtained from equation (2), $z_0=0$. The last term in
the integration gives the probability that particles collide at a given point z:  $\Pi(b)=1-e^{z_0-z/\lambda}$ , and the 
mean free path $\lambda=\frac{1}{\sigma \rho}$.
We can further simplify the integration by assuming that the particles collide after traveling a distance $z_0$, i.e. z=0.
The estimate is compared to experimental data from \cite{wong} in figure 2  and we can see that 
we obtain rather surprisingly a reasonable agreement.  In particular, the agreement is improved if one takes into account
the collision probability in eq.(2) (open squares in fig.2).  
\begin{figure}[ht]
\centerline{\hspace{-0.5in}
\includegraphics[width=6.5in,height=4.0in,angle=0]{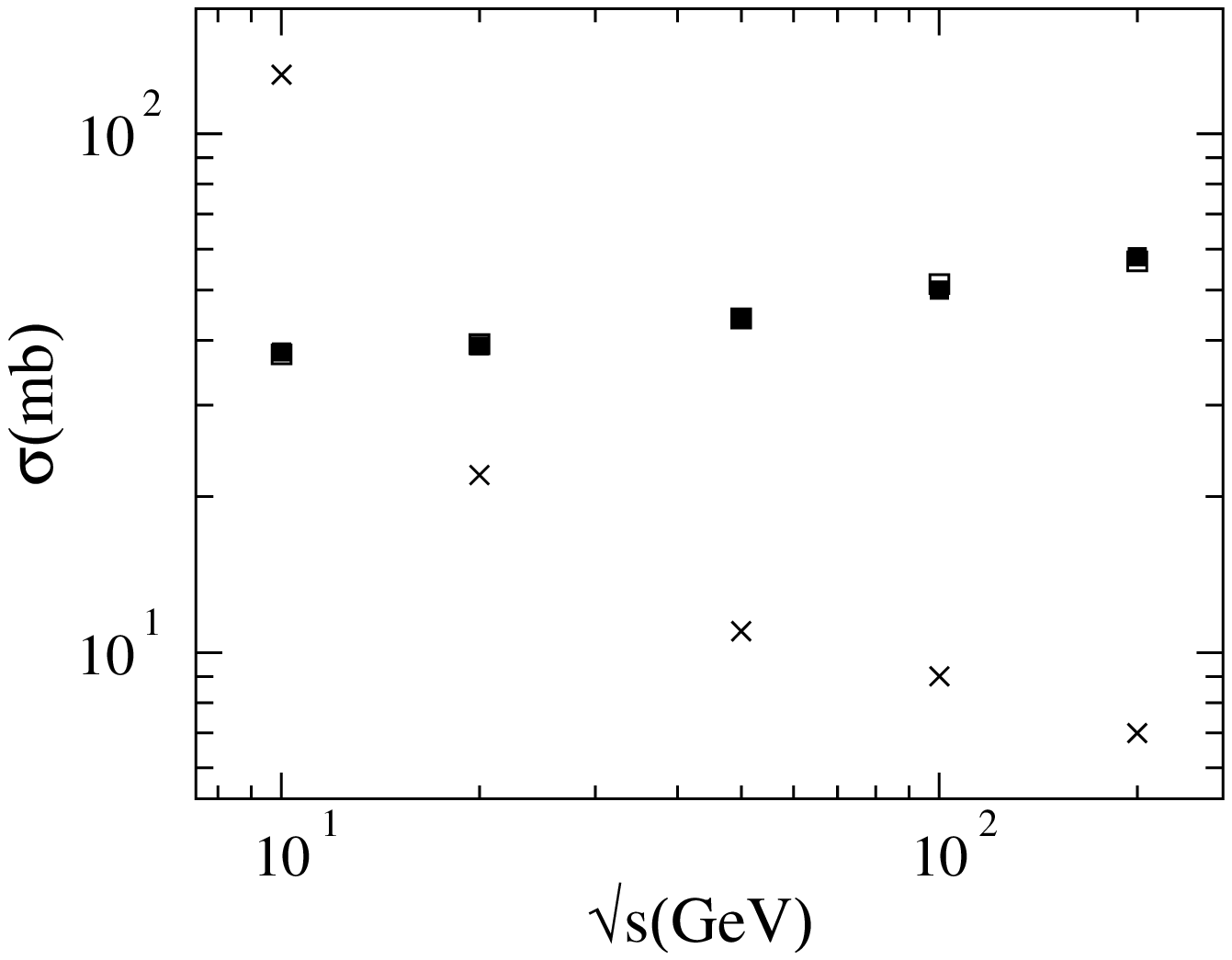}}
\vspace{0.2in} \caption{Cross section versus energy from the analytical estimate, eq(5).  The data are given by the full squares, open squares are our  results 
while the crosses are the input values in the probability,eq.(4) .}
\label{fig3}
\end{figure}

In our simulation a collision between particles is probabilistic, different from other approaches\cite{cug,bert,aldo,sa1} where
a geometrical  assumption is made for collisions (black sphere).  This implies that collisions (both elastically and inelastically),
can occur  as well for impact parameters larger than $\sqrt(\sigma/\pi)$.  Thus in order to know what is the effective cross section implemented
we need to integrate over impact parameters.  For instance, for inelastic collisions we get:
    \begin{equation}
\sigma_i=2\pi \int ^{b_{max}}_0b\Pi(b)db
\label{cs2}
\end{equation}
Notice that this equation for $\sigma$ contains a cross section also in the probability, in some sense we are transforming a
sharp cutoff probability  into a diffuse one.  Recall that  the inelastic process depends from the reaching of a critical
energy density in our model, thus from $b_{max}$, and this process will result in a certain dependence on energy of the
cross section.  Our strategy is to fit the input cross section in the probability in order to reproduce the experimental values.  This is
shown in figure (3) where we plot the experimental cross sections (full squares) versus energy\cite{wong}. We reproduce rather well the data (open
 squares) when we choose an input cross section given by the crosses.  We see that we are able to reproduce the results especially at high
 energies.  The input cross section is one of the free parameters of our model together with the critical energy density and the quarks masses.  Once
 we have fixed such values we obtain the number of charged particles shown in figure(2).
In the numerical simulation that we discuss below, 
we do not make any assumption and the trajectory is followed in time without simplifications. It is also included the possibility that partons are created
from secondary collisions during the expansion phase. For our purposes,
the result found here gives a strong hint that the assumption we make are rather reasonable.  However, we expected
this result since we know that statistical models work well already in pp and ee collisions as claimed for instance in
\cite{bec}.  

\section{Numerical simulations}

Numerically the collisions are followed in time for given initial conditions and beam energies\cite{tan,aldo}.  We have
performed simulations from tens of GeV initial energy to several TeV which should be soon experimentally available at 
CERN.  The average number of produced charged particles and the corresponding cross sections are calculated by 
generating many events for each energy and for many impact parameters b.  The number of events is generated 
proportionally to  the impact parameter b as usual, eq.(3).  In fig. (4)  we plot the obtained cross section as function of
$\sqrt(s)$.  A parametrization of the data \cite{wong} is given by the full dots and our numerical results by the open dots.
The experimental points at the highest energy were obtained in air shower experiments \cite{hon}.

\begin{figure}[ht]
\centerline{\hspace{-0.5in}
\includegraphics[width=6.5in,height=4.0in,angle=0]{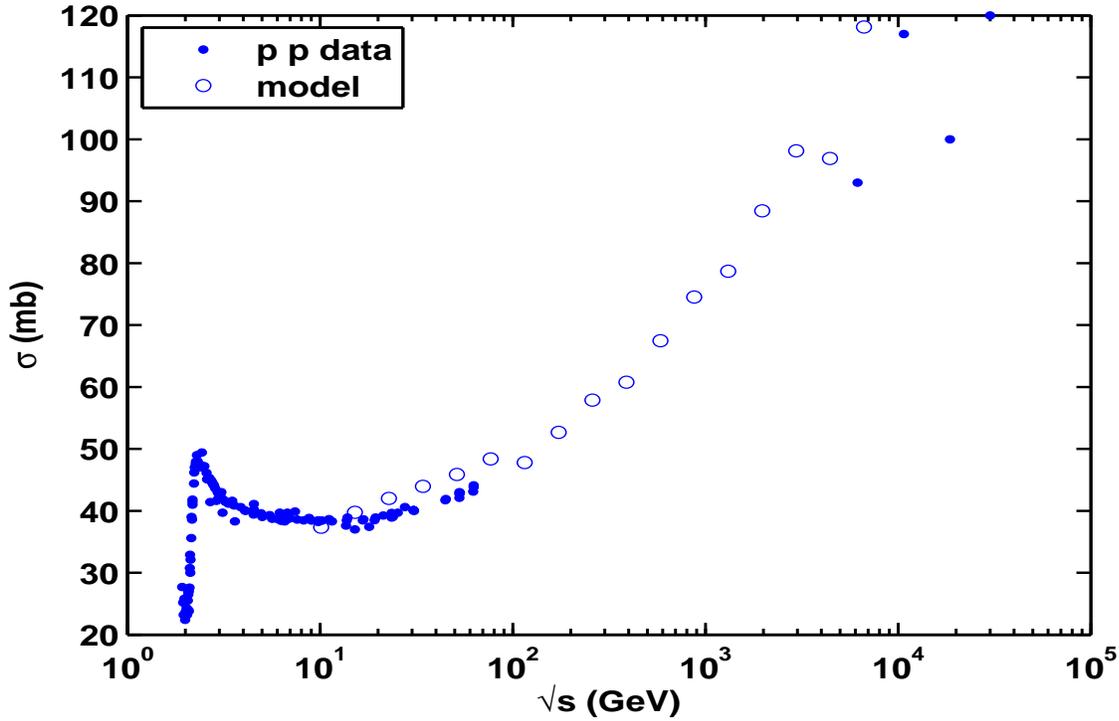}}
\vspace{0.2in} \caption{Cross section versus energy.  The data is given by the full dots, open dots are our numerical results.}
\label{fig4}
\end{figure}

The average number of  produced charged particles can be calculated similarly to  the previous section and the
results are given in fig.5.
\begin{figure}[ht]
\centerline{\hspace{-0.5in}
\includegraphics[width=6.5in,height=4.0in,angle=0]{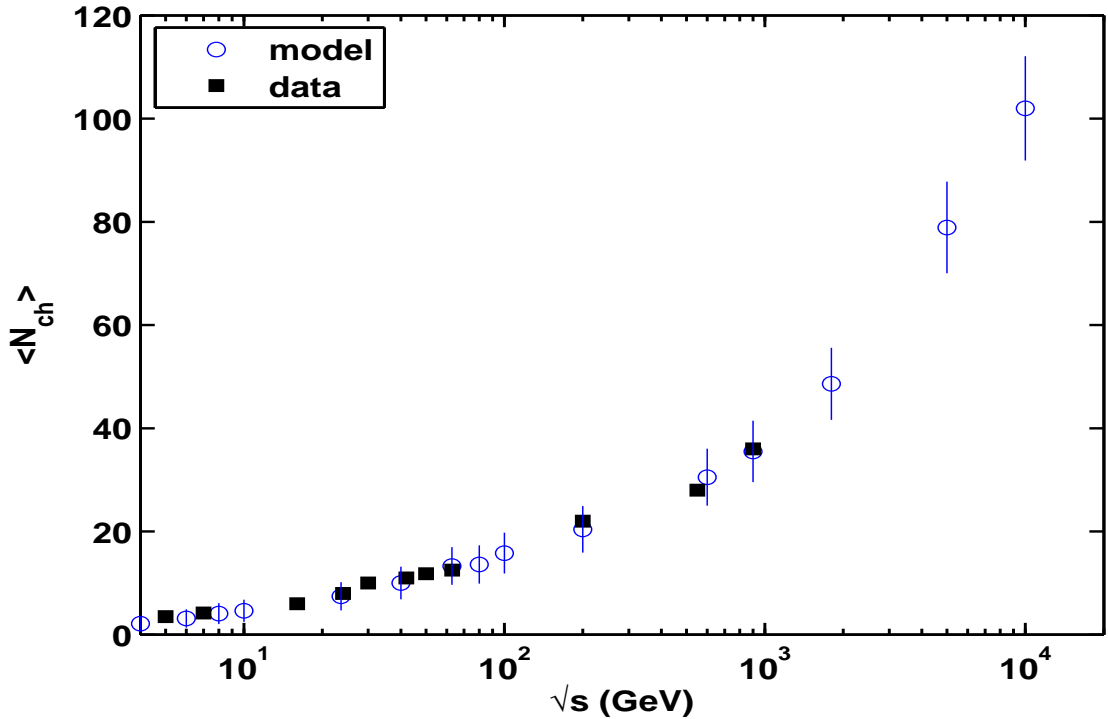}}
\vspace{0.2in} \caption{Number of produced charged particles versus energy.  The data is given by the full symbols.}
\label{fig5}
\end{figure}

Because of the probabilistic nature of our approach the number of particles produced depend strongly on the
 impact parameter b. In figure (6) we plot the number of produced charged particles versus $b/b_{max}$ at
different beam energies. Since the maximum impact parameter increases with energy, see eq.(2), we reproduce
the well known Blacker, Edger and Larger (BEL) effect\cite{gei}: for higher energies the elementary cross section increases thus the
number of involved impact parameters increases as well.
\begin{figure}[ht]
\centerline{\hspace{-0.5in}
\includegraphics[width=6.5in,height=4.0in,angle=0]{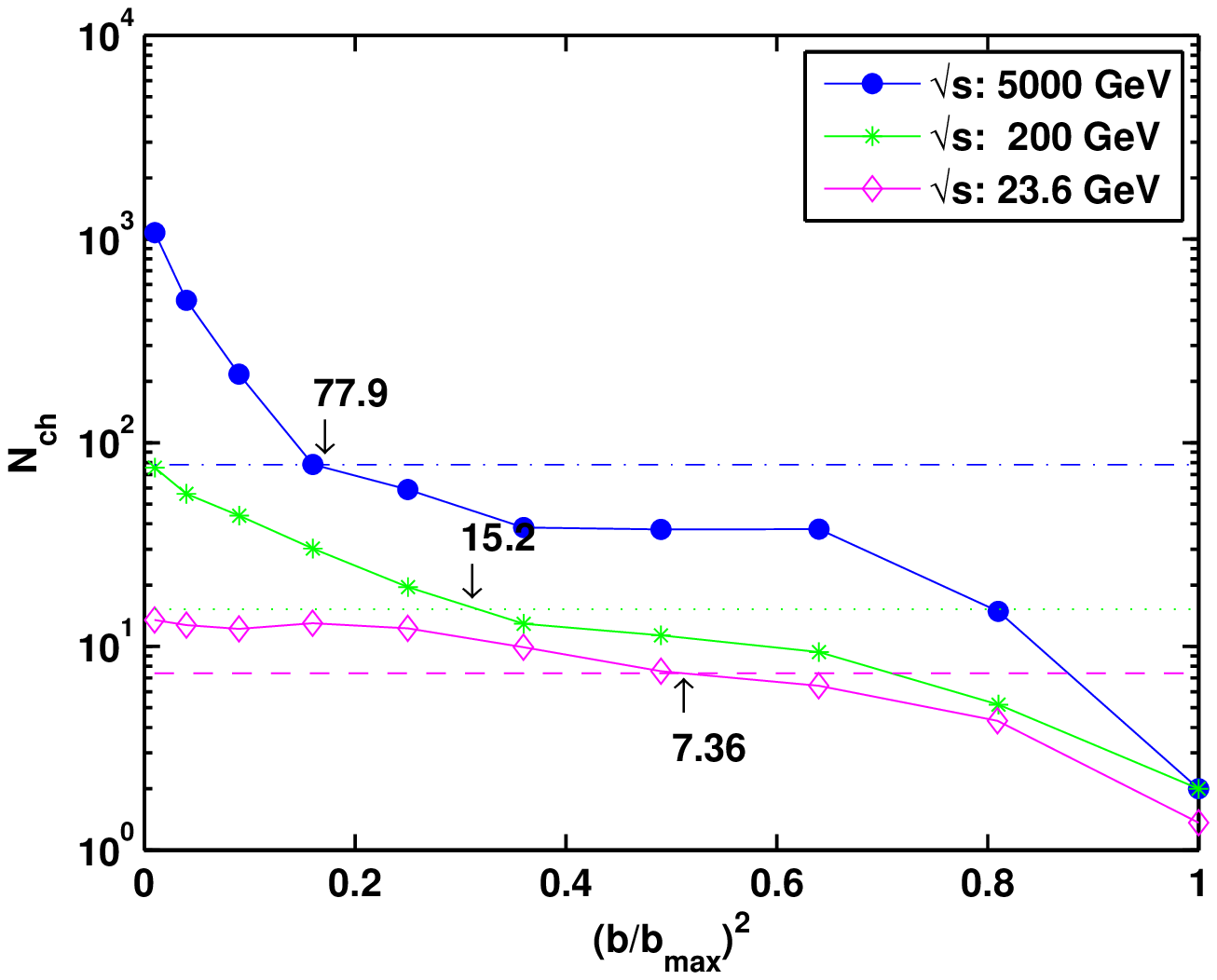}}
\vspace{0.2in} \caption{Number of produced charged particles versus b at different energies.  The dashed lines give the average number 
(over impact parameters) of produced charged particles.}
\label{fig6}
\end{figure}

A more sensitive quantity to obtain is for instance the pseudorapidity distribution of charged particles.  In our approach
 after the partons are created they are given an initial random momentum along the z-axis (the beam axis) and proportional to the
 initial momenta of the colliding hadrons.  Total energy and momenta conservation are enforced.  In the following time
 steps, the partons expand and they could collide with other particles getting in this way some momentum
 in the transverse direction to the beam axis.  Also diquarks can be formed with a given probability.  The diquarks can
 subsequently collide again and form a hadron.  We  fit the parton-particle and diquark-particle cross sections in order to 
 reproduce the experimental data on pseudorapidity distributions \cite{wong,aln}.  In fig.7 the distribution is plotted at
 $\sqrt(s)=200 GeV$ for different  parton-particle cross sections.
   Notice that the transverse momentum is obtained
 \begin{figure}[ht]
\centerline{\hspace{-0.5in}
\includegraphics[width=6.5in,height=4.0in,angle=0]{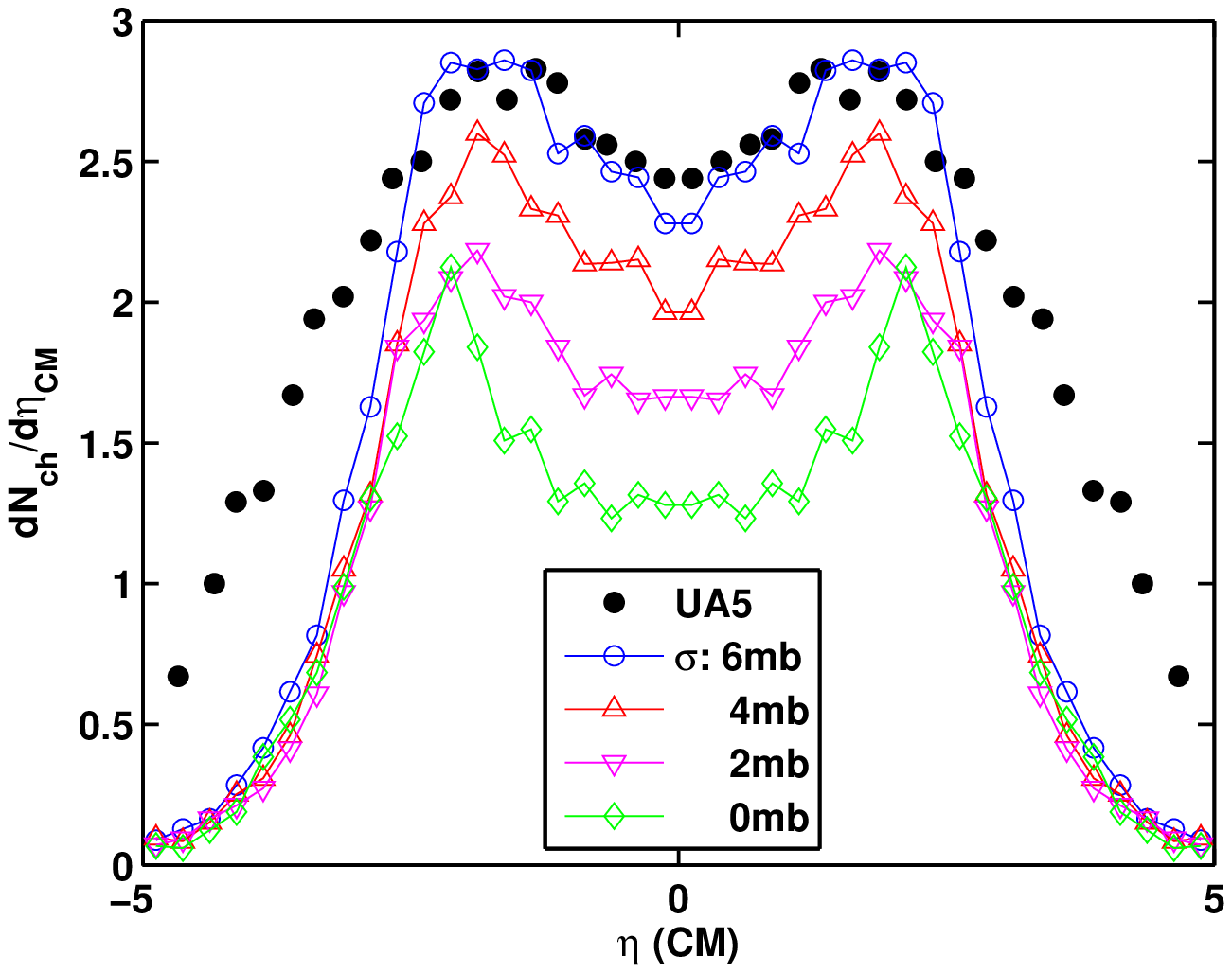}}
\vspace{0.2in} \caption{Pseudorapidity distribution at $\sqrt s=200GeV$, the data points are given by the full dots\cite{aln}. Theoretical calculations
are given by the open symbols for different parton-particle cross sections. }
\label{fig7}
\end{figure}
 not only from these collisions but also from the coalescence and further decay of resonances later on when the
 density is smaller than a critical value as obtained from the MIT bag model.  However, we notice that a cross
 section of few $ mb$ works rather well.  In figure 8 we plot the distributions for different initial energies.
 The agreement to data is fair.
 
\begin{figure}[ht]
\centerline{\hspace{-0.5in}
\includegraphics[width=6.5in,height=4.0in,angle=0]{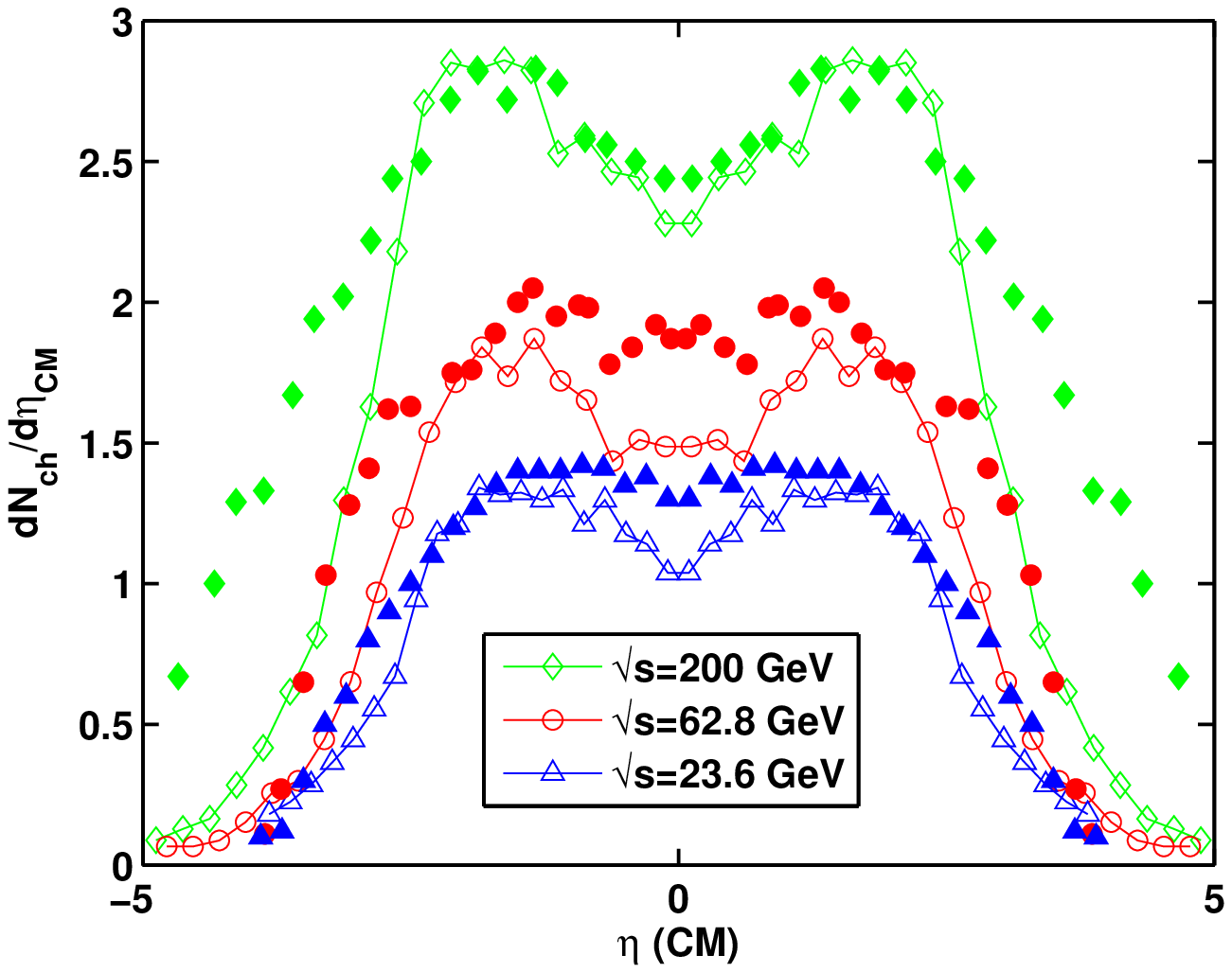}}
\vspace{0.2in} \caption{Pseudorapidity distributions at different beam energies.  Data are given by the full symbols\cite{aln}.}
\label{fig8}
\end{figure}

 In fig. (9) we have performed calculations at energies larger than  1 TeV  which is our prediction
 for some future CERN experiments.

 \begin{figure}[ht]
\centerline{\hspace{-0.5in}
\includegraphics[width=6.5in,height=4.0in,angle=0]{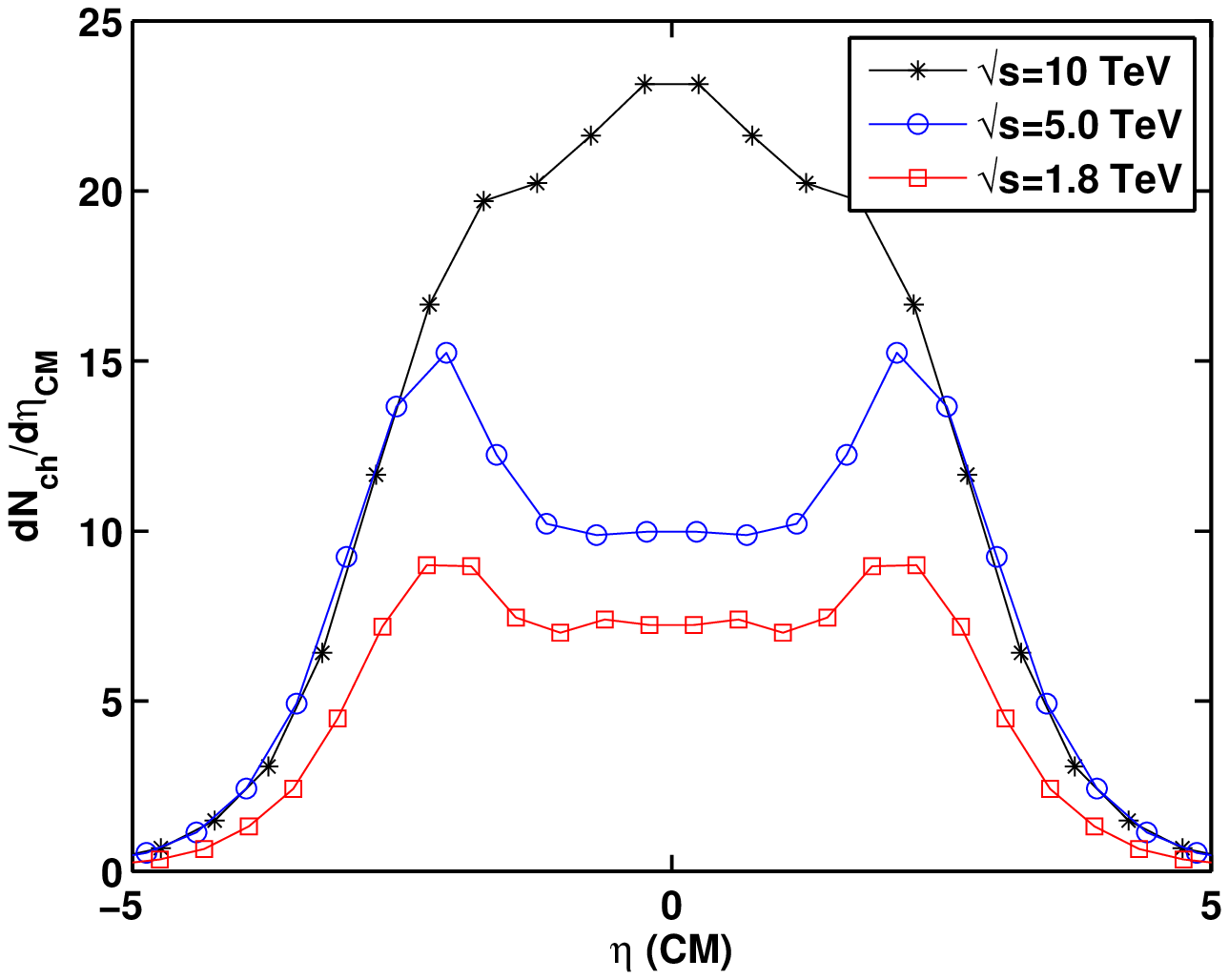}}
\vspace{0.2in} \caption{Pseudorapidity distributions at 1.8,5 and 10 TeV.  }
\label{fig9}
\end{figure}

\section{Mass Distributions}

The model we have proposed makes use of a coalescence mechanism to produce the new hadrons.  The physics
we have in mind is that after the pp collision, partons are created in a very dense region.  The partons expand and
eventually collide thus getting some transverse momentum.  When the local density decreases below a certain value,
partons coalesce into resonances.  Those hadrons expand further 
and eventually decay or collide with other particles.  This 
mechanism is reminiscent of droplets formation at lower energies for a liquid to gas phase transition\cite{bon00}.  In the same
sense we could compare the partons to the vapor and the hadrons to the liquid.  In the expansion the system crosses
from one phase to the other.  In the crossing we expect fluctuations to be large which could be 
reveled from the fluctuations of momenta distributions of hadrons\cite{terr05}.  A strong signature which appears also 
in non equilibrium cases and for a reduced number of particles is a power law in the mass distributions \cite{bon00}.  
In a pp collisions we have seen that the multiplicity of produced particles could be large, thus we could expect in
the model that a self-similar mass distribution appears.  In fig.(10) we plot the numerical mass distributions $Y(> m)$ which
is the number of fragments obtained with mass greater than m\cite{tur}.  We have adopted this definition rather than
the commonly used yield to avoid numerical fluctuations in the simulations.  We can fit  such a distribution with
the expression: $Y(>m)=cost\times m^{-\tau+1}$.  From the fit we obtain a value of $\tau=2.3$ which is very 
reminiscent of the Fisher law for a gas to liquid phase transition.  Thus this result hints for a second order phase transition.  
As we see from the figure, the power law is rather strong and independent on the beam energy, the rapid falloff
at around 2 GeV is simply due to the inclusion  in the calculations of resonances up to 2 GeV and to the low statistics.
\begin{figure}[ht]
\centerline{\hspace{-0.5in}
\includegraphics[width=6.5in,height=4.0in,angle=0]{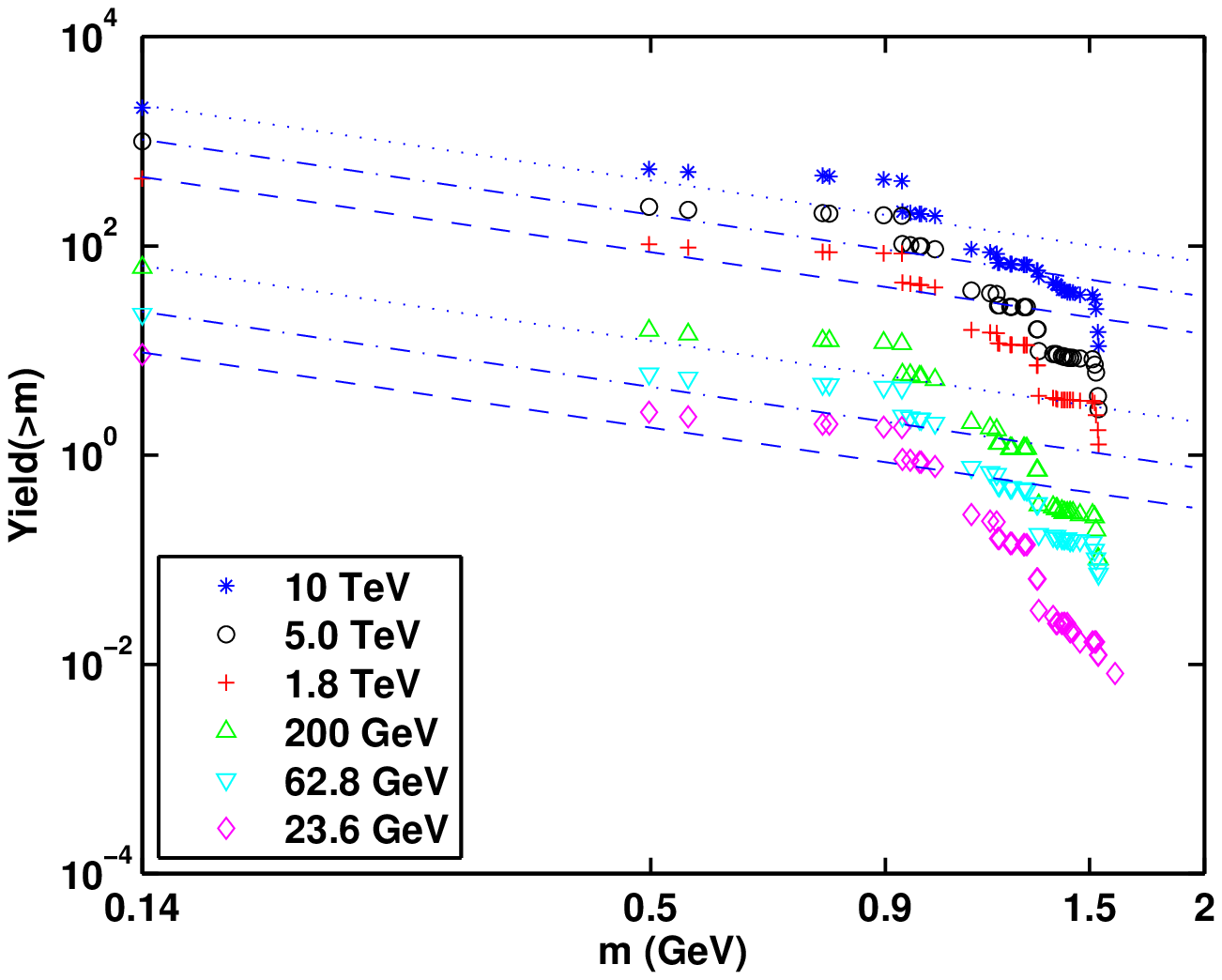}}
\vspace{0.7in} \caption{Mass distributions at different energies. 
The lines are power law fits.}
\label{fig10}
\end{figure}

We expect that this result is independent on the way the plasma was formed, thus should show even better in nucleus-
 nucleus collisions.  We have collected some experimental data for Au+Au collisions at RHIC \cite{star} which are displayed in fig.(11).
 In this case since the data has enough statistics we have plotted directly the measured yield.  Also notice that now the power law is extended
 to masses larger than 2 GeV.  The bumps and dips in the figure are due to  detected resonances of different widths.  This is somehow similar
 to the corresponding oscillations in the mass distributions at lower energy for liquid-gas phase transition due to shell effects\cite{bon00}
  \begin{figure}[ht]
\centerline{\hspace{-0.5in}
\includegraphics[width=6.5in,height=4.0in,angle=0]{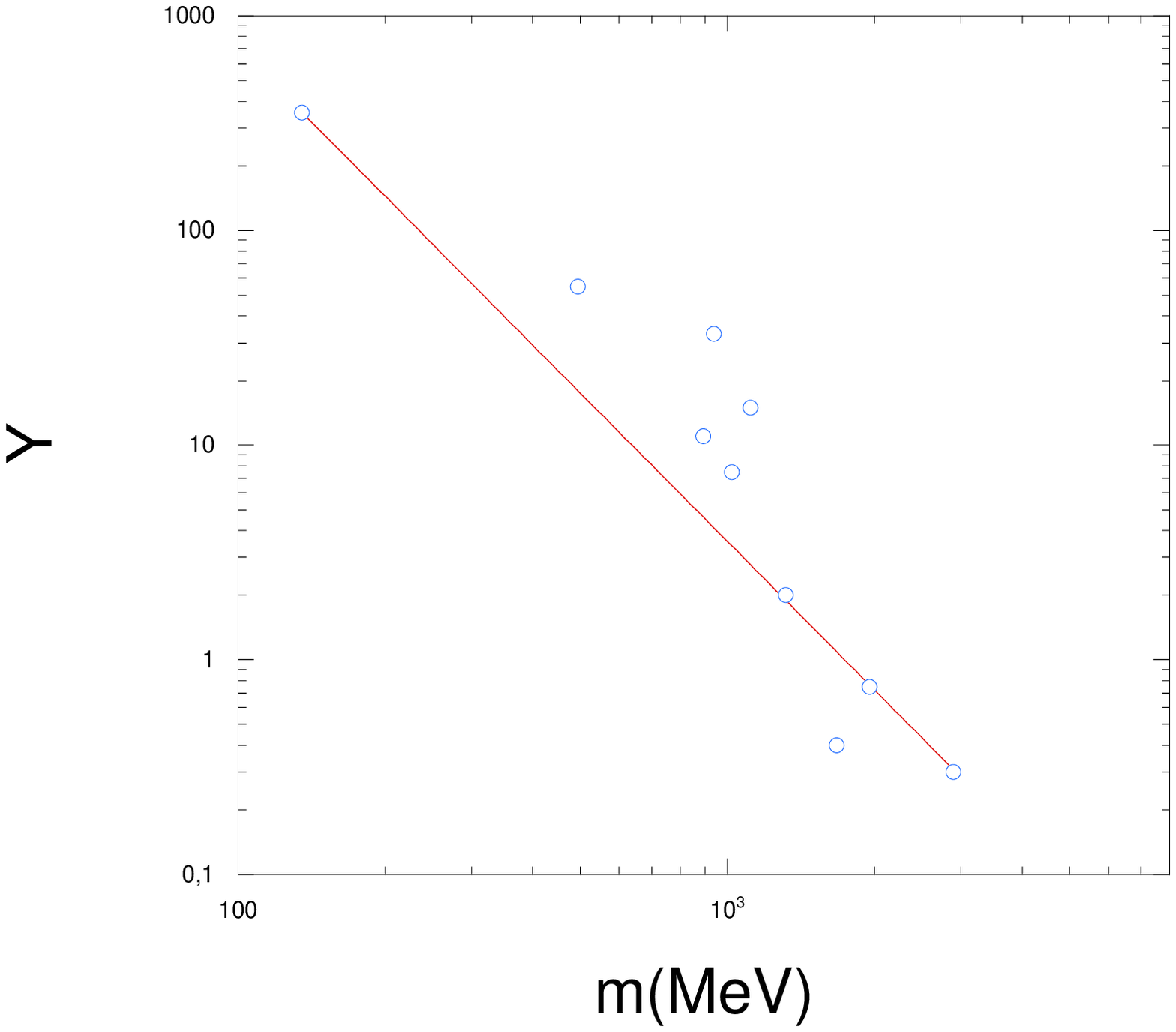}}
\vspace{0.7in} \caption{Mass distributions at 200GeV in Au+Au collisions\cite{star}. 
The line is a power law fit.}
\label{fig11}
\end{figure}

 \section{conclusions}

In this work we have applied a recently introduced transport approach to
study proton-proton collisions.  This is a necessary step if we want to simulate nucleus-nucleus collisions.
The model includes  resonance formation and  their decays.  The possibility of a QGP
is included based on the MIT bag model as well as quantum statistics.  We have seen that we can reasonably reproduce available data on pp
collisions and also be able to make prediction at LHC energies.  The main assumption, we believe, is inspired by thermal models which we have
generalized  to non-equilibrium situations.  The success in the data reproduction suggests that the system is chaotic enough to justify our assumption.
Pseudorapidity distributions can be reproduced assuming some suitable parton-parton collision cross section.  
  Hadronization occurs via partons coalescence which is similar to coalescence of vapor into drops.  We have explored this analogy by analyzing
   physical quantities which are relevant in phase transitions such as mass distributions and found a power law with an exponent compatible with a
   Fisher's law analysis.  This is seen not only in our calculations but also in Au+Au experimental data.  This signature is at the moment not sufficient to pin
   down the phase transition and should be considered as a $\it curious$ result.  In following works we will present a detailed analysis for a critical behavior and we hope
   to see soon corresponding data to confirm or reject our scenario.

 Z.G.T  acknowledges the financial support from  INFN  (where most of
the work was performed).  We also thank V.Greco for stimulating discussions and comments.


\begin{thebibliography}{99}
\bibitem{tan}Z.G.Tan and A.Bonasera, Nucl.Phys. A784(2007)368.
Z.G.Tan, D.M.Zhou, S.Terranova and A.Bonasera,  nucl-th/0606055, and proc. XXII Winter Workshop on
Nuclear Dynamics, pag.31, W.Bauer et al. eds.(2006).
\bibitem{karsch}F. Kartsch, Nucl.Phys. A698, 199c(2002).
\bibitem{gei}
K. Geiger and B. M$\ddot{u}$ller, Nucl. Phys. B 369, 600 (1992);
K. Geiger, Phys. Rep. 258, 237 (1995).
\bibitem{cug}
J. Cugnon, T. Mitzulani, J. Meulen, Nucl. Phys. A 352, 505 (1981).
\bibitem{bert}
G. F. Bertsch and S. Das Gupta, Phys. Rep. 160, 189 (1988).
\bibitem{sa1}
Ben-Hao Sa, Wang Rui-Hong, Zhang Xiao-Ze, Zheng Yu-Ming, and Lu Zhong-Dao,
Phys. Rev. C 50, 2614 (1994).
\bibitem{aldo}
A. Bonasera, F. Gulminelli, and J. Molitoris, Phys. Rep. 243,1(1994); A.Bonasera, T.Maruyama Progr.Theor.Phys. 90(1993)12.
\bibitem{daimei} D.M.Zhou, S.Terranova and A.Bonasera,  nucl-th/0501083.
\bibitem{las}
L. P. Csernai, "Introduction to relativistic heavy ion
collisions", John Wiley and Sons, 1994.
\bibitem{wong}
 C.~Y.~Wong, {\it Introduction to High-Energy Heavy ion Collisions},
 World Scientific Co., Singapore,1994.
\bibitem{bert1}G. M. Welke and G. F. Bertsch, Phys. Rev. C 45, 1403 (1992).
\bibitem{sa04} Ben Hao Sa and A. Bonasera, Phys. Rev. C 70, 034904 (2004).
\bibitem{hagel}K.Hagel et al. PRC62(2000)034607; A.Bonasera et al.PLB244(1990)169.
\bibitem{dorso}C.Dorso et al., PRC60(1999)034604; M.Belkacem et al., PRC54(1996)2435.
\bibitem{enzo}V.Greco et al., PLB595(2004)202.
\bibitem{bec}F.Becattini and U.Heinz, Zeit. f. Phys.C76(1997)269.
\bibitem{hon} M.Honda et al. PRL70(1993)525.
\bibitem{aln}G.J.Alner et al., UA5 coll., Z.Phys.C33(1986)1.
\bibitem{bon00} A.Bonasera et al., Rivista Nuovo Cimento, 23(2000)1.
\bibitem{terr05} S.Terranova, D.M. Zhou and A.Bonasera,
nucl-th/0412031, EPJA26, 333 (2005).
\bibitem{tur}D.L.Turcotte, Fractals and chaos in geology and geophysics, Cambridge university press(1992).
\bibitem{star}G. Van Buren, Star collaboration, arXiv:nucl-ex/0211021v3 (2002).
\end{thebibliography}
\end{document}